\documentclass[12pt]{article}

\usepackage{graphicx}
\usepackage{color}
\usepackage{mathptmx}      % use Times fonts if available on your TeX system
\usepackage{amssymb}
\usepackage{amsmath}
\usepackage{ifsym}
\usepackage{sidecap}
\usepackage[colorlinks, linkcolor=blue, citecolor=blue, urlcolor=blue]{hyperref}

\usepackage{amssymb}
\usepackage{lscape}

\begin{document}

\noindent
\medskip
\centerline
{\Large\bf On the possibility of laboratory simulation}

\noindent
\medskip
\centerline
{\Large\bf  of quasi-spherical accretion onto black holes}

\noindent
\medskip
\centerline
{\Large\bf  with a shallow-water experimental setup}

\bigskip
\bigskip

\noindent
%\centerline
{\large A.\,V.\,Semyannikov\footnote{\sl E-mail address: semyannikov.alexander@volsu.ru},  I.\,G.\,Kovalenko\footnote{\sl E-mail address: ilya.g.kovalenko@gmail.com}
}

\bigskip
\noindent
{\it Volgograd State University,
         Universitetskij Prospekt, 100,  Volgograd 400062, Russia
}

\begin{abstract}
We describe the concept of a shallow-water setup for simulation of gas accretion onto a black hole in the mode of a quasi-spherical accretion. The bottom for the shallow-water container must have the funnel-shaped curvilinear concavo-convex shape. We calculate the configuration surface of the properly shaped bottom that simulates precisely the Newtonian or pseudo-Newtonian gravitational potentials. Like the spatial part of the Schwarzchild metric, the funnel's surface metric has a (removable) singularity at the finite distance from the funnel's center and places the certain funnel's depth which we call `gravitational length'. The gravitational length is analogous to the gravitational radius and defines the equivalent of  the black hole's mass in the laboratory model.  The mass equivalent corresponds to $\sim 0.367\cdot 10^{12}$ g for the funnel as deep as 5 cm. We define more precisely the inviscid shallow water equations for the arbitrary bottom curvature. We show that in general case the shallow water pressure obeys the non-barotropic equation of state. We suggest the schematic course for experiments for simulation of accretion in a thick accretion disk mode as well as the Bondi-Hoyle accretion.
\end{abstract}

%\bigskip\noindent
%{\it Keywords}
%%% keywords here, in the form: keyword \sep keyword
%

%%%% PACS numbers
%\PACS{97.10.Gz, 97.60.Lf, 01.50.Pa}

\section{Introduction}

The laboratory shallow-water simulation of astrophysical hydrodynamic flows is of deep and rich history \cite{NezSnezh90,Dolzh90}. As a rule, the point at issue is simulation of rotating fluids in which both the  Coriolis force and shear due to differential rotation play the key role in structure formation \cite{NezSnezh90,Antipov83,Morozov84,Chernikov12}. Among these are accretion flows realized in the mode of disk accretion, such as a thin Keplerian disk in which the non-radial velocity is much greater than the radial one.

Meanwhile the other accretion modes exist in which rotation plays the secondary role but they can be simulated with shallow-water tanks as well. These include (i) the spherical accretion (non-radial velocity components are vanishingly small compared to the radial component) \cite{Bondi52, Das07}; (ii) thick accretion disk (non-radial velocity components are comparable with the radial component) \cite{Chakra89}. Simulation of gravitational metric in proximity of the black hole through the use of shallow-water experiments but not the accretion flow itself constitutes the separate line of modeling \cite{Unru05}. In our opinion, this line of research has a great potential and we will touch on the theme in a subsequent paper. Here we explain how the shallow-water setup can be used for simulation in accordance with the items (i)-(ii) which we jointly name the models of quasi-spherical accretion.

The cumulation of perturbations in converging flow is a more significant effect in accretion with low angular momentum \cite{KovEre98}. Given that the degree of cumulation is defined as the outer to inner radius ratio one cannot expect the cumulation ratio greater than two orders of magnitude in the experiment. Even then the effects of amplification of perturbations  can be very much in evidence \cite{Japan14} and the laboratory modeling makes sense.

The suchlike setup is already constructed \cite{Foglizzo12}. It is used for simulation of instability of standing shock wave in an accreting flow of pre-supernova core collapse. However, the approach used in  \cite{Foglizzo12} cannot be considered as the wholly satisfactory one. The sink in experiments \cite{Foglizzo12} in which the shallow water layer flows down to outlet in the bottom has the shape which does not obey the key requirement for the laboratory model. This requirement from our point of view should be as follows:

{\it the equation of motion of the point particle sliding down the funnel's surface with no angular momentum must duplicate the equation of motion of the body freely falling onto a black hole.}

For the Schwarzchild metric one can consider as a case in point the integral of gravity body dropping from infinity with zero initial velocity. This integral coincides with the energy integral of a free falling body in the Newtonian gravity field \cite{LandLifsh}
\begin{equation}\label{Int_energy_Sch}
  U_R^2/(2c^2) - r_g/(2R) = 0.
\end{equation}
Here $R$  is the spherical radius, $U_R$, the physical radial velocity of the particle,  $c$, the speed of light, $r_g$, the gravitational radius of a black hole. A similar integral corresponding to the profile of the funnel in \cite{Foglizzo12} will have a qualitatively different kind, in which the kinetic energy contains an explicit strong dependence on the cylindrical radius $r$
\begin{equation}\label{Int_energy_Fogl}
 \left( 1+ \frac{H^2}{r^2} \right) \frac{U_r^2}{2 v_{II}^2} - \frac{H}{2r} = 0.
\end{equation}
Here $U_r$ is the cylindrical radial velocity, $H=5.6$ cm, $v_{II}=(2gH)^{1/2}$. Furthermore, the shallow-water equations presented in \cite{Foglizzo12} correspond to liquid flow along the surfaces of slowly varying curvature and do not consider the change of weight of the liquid column due to centrifugal forces. The impact of latter ones should be significant near the hole in the narrowest part of the funnel, through which the liquid leaves the funnel.

Perhaps, for realization of the experiment with the shock wave instability these differences are not so important, but when setting more refined experiments they can make substantial quantitative or qualitative adjustments.

We give a detailed derivation of the funnel surface geometry for simulation of gas accretion in the Newtonian or pseudo-Newtonian potential corresponding to the above request, as well as the derivation of equations for shallow water on a curved surface. We show that the metric of funnel surface is similar to the spatial part of the Schwarzschild metric in the sense that it is characterized by a certain scale, similar to gravitational radius, on which the metric has a removable singularity.

\section{Description of laboratory setup with shallow water. Flow phenomena
}

In the real laboratory shallow water facilities the fluid is essentially dissipative due to the bottom friction. It is reasonable to make the detailed estimation of the influence of friction at the final stage of the work when adjustment of the funnel profile for the flow contour of particular interest will be required. While we are at the stage of development of the concept, we consider ourselves justified in confining ourselves to the approximation of non-viscous fluid.

\begin{figure}
\includegraphics[width=\columnwidth]{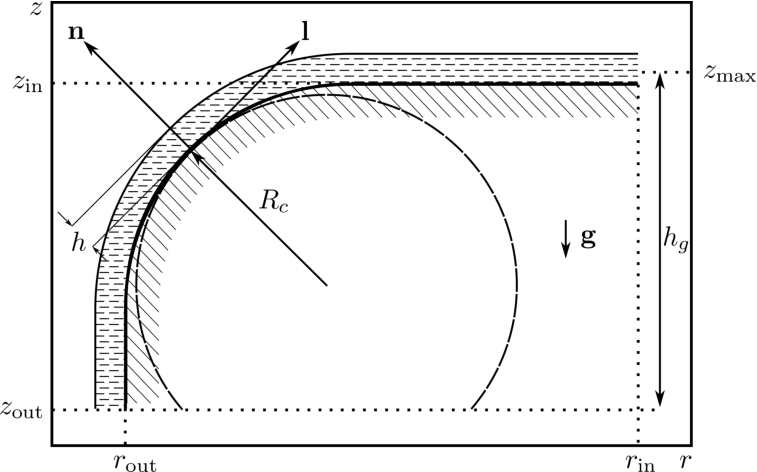}
\caption{\label{fig:1}
The profile of device with shallow water. A vertical sectional view of an axisymmetric funnel in the cylindrical coordinate system is shown. Explanation of symbols are provided in the text.}
\end{figure}

We believe that the setup must have the shape schematically shown in Fig.~\ref{fig:1}. The bottom for thin liquid layer of thickness $h$, flowing downward in the gravity field, is the axisymmetric funnel installed in vertical orientation. Its profile is defined as the rotational surface $r=r_0(z)$, where $z$ is the coordinate along the vertical axis. The thickness $h$ is defined along the normal to the bottom. Liquid is supplied at the outer (upper) boundary of the funnel at $(r_{in}=r_0(z_{in}), z_{in})$, and leaves the funnel from below through a hole in its bottom level $ (r_{out}=r_0(z_{out}),  z_{out})$. Fluid may be supplied either strictly in a radial axis, or with a twist in the azimuthal direction. If necessary, a funnel can run on an axis.

The reduction to two-dimensional description is as follows. Ad initium the flow of an incompressible fluid is considered to be three-dimensional in the Euclidean space $E_3$ with the metric tensor $e_{ik}$ written in the cylindrical coordinate system as $e_{ik} = {\rm diag}(1, r^2, 1)$. It is convenient to define the bottom profile parametrically as $(r_0(l), z_0(l))$ in the natural orthogonal coordinate system $(x^1=l, x^2=\varphi, x^3=n)$, where $l$ is the pseudo-radius, the coordinate along the generatrix of the funnel. We define $l$ as a natural parameter of generatrix, $\varphi$ stands for the azimuth angle, $n$ is the coordinate normal to the bottom (Fig.~\ref{fig:1}.). The grid is adjusted so that the value $n=0$ corresponds to the bottom. We denote the tensor $e_{ik}$ written in new coordinates $(l, \varphi, n)$ by $g_{ik}$.

After the standard procedure of averaging over transverse to the layer of fluid coordinate $n$ we come to the fluid flow description in a curved two-dimensional $V_2$, $V_2 \subset E_3$ with coordinates $(x^1=l,\  x^2=\varphi)$ and  metric $g_{\alpha\beta}$ whose matrix elements are the elements of $g_{ik}$ at $x^3=0$. Here and below, the prime marks the derivatives with respect to a single argument  $l$, the Greek indices refer to the summation from 1 to 2, and the Latin indices run from 1 to 3.

Given that $g_{ll}(n=0) = e_{rr}\left( r_0^{\prime}\right)^2 + e_{zz}\left( z_0^{\prime}\right)^2 = 1$, we have
\begin{equation}\label{g11_g22}
g_{\alpha\beta}  = {\rm diag}(1, r_0^2(l)), \qquad
\partial_l g_{\varphi\varphi}  = 2r_0 r_0^{\prime}.
\end{equation}

The coordinate n is selected so that at least in an infinitesimal neighborhood of $n=0$, it acts as a natural parameter. This leads to the equations for $n=0$
\begin{equation}\label{g33}
g_{nn} = 1, \quad
\partial_l g_{nn} = 0, \quad
\partial_n g_{\varphi\varphi} = -2r_0\sqrt{1-(r_0^{\prime})^2}.
\end{equation}
The latter one follows from the fact that the square of the radius gradient is invariant with respect to rotations, and which implies $(\partial_l r)^2+(\partial_n r)^2 = (\partial_r r)^2+(\partial_z r)^2 =1$. Finally, after a simple differential geometry analysis we find for $n=0$
\begin{equation}\label{g11_33_diff}
\partial_n g_{ll} = \theta_{-}^{+}\cdot {2}/{R_c},
\end{equation}
where $R_c$ is the local radius of curvature of the surface $n=0$ at $l$, which we define as a positively determined in accordance with the formula
\begin{equation}\label{R_c}
R^{-1}_c(l) =
(r_0^{\prime})^2\left| \left( \sqrt{1-(r_0^{\prime})^2}/ r_0^{\prime} \right)^{\prime} \right| .
\end{equation}
We prefer to hide the sign of curvature in the mnemonic factor $\theta_{-}^{+}$, which is equal to $+1$ if the funnel surface is convex upwards, or to $-1$ if the surface is concave. The derivative $\partial_l g_{nn}(n=0)$ does not require the explicit calculation, as the procedure of averaging over normal coordinate eliminates the term in the hydrodynamic equations containing it.

The hydrodynamic equations of an incompressible fluid in the covariant $(3+1)$-form are written as
\begin{equation}\label{EqHydroGenCont}
\rho \nabla_{;j} u^{j} = 0 .
\end{equation}
\begin{equation}\label{EqHydroGenEuler}
\frac{\partial u^i}{\partial t} + u^{j} \nabla_{;j} u^{i} = - \frac{1}{\rho} \nabla^{,i} p - \nabla^{,i} \phi  ,
\end{equation}
Here $u^i$ is the geometric 3-velocity, $p$, the pressure, density $\rho$ is assumed to be a time-independent constant, $\phi$ is the potential of the uniform gravity field, $\phi = gz +{\rm const}$, where $g$ is the free fall acceleration, the characters $`,'$ and $`;'$ designate partial and covariant differentiation, respectively.

In the notation of the physical 3-velocity $U_i$, the components of which are prescribed as
\begin{equation}\label{uU}
 U_{j} = u^j H_j ,
\end{equation}
where $H_{i} = \sqrt{|g_{ii}|}$ are the Lame coefficients, the equations of fluid motion can be expressed explicitly as follows:
\begin{equation}\label{EqHydroPhysCont}
\rho \partial_i  \left( U_{i} \sqrt{ \det( g_{ik} )}/H_{i} \right) =0,
\end{equation}
\begin{equation}\label{EqHydroPhysEuler}
\begin{split}
 \left(\partial_t  + \frac{ U_{j} }{H_{j}} \partial_j \right)  U_{i} & + \frac{ U_{j} }{H_{j}} \left( \Gamma^{i}_{k j} \frac{H_{i}}{H_{k}} U_{k} - \Gamma^{i}_{i j} U_{i} \right)  = \\
 = & - {H_{i}e^{ik}} \left( \partial_k p + \rho \partial_k \phi \right)/{\rho} ,
\end{split}
\end{equation}
where $\Gamma^{i}_{kj}$ are the Christoffel symbols.

We integrate equations \eqref{EqHydroPhysCont}-\eqref{EqHydroPhysEuler} along the coordinate $x^3$ over the liquid layer height $h=h(t, x^1, x^2)$) and make use of approximations which are considered to be common for the shallow water theory: (i) the longitudinal (along the bottom) disturbances in the fluid have a length $\lambda$ corresponding to the bounded interval of scales $R_c\gg \lambda \gg h$ that allow us to use the assumption that at every instant in each liquid column with coordinates ($x^1, x^2$) the flow rapidly relaxes to the hydrostatic equilibrium; this means that the inertia of the liquid transverse movements (normal to the bottom) can be ignored; (ii) as a result, the transverse velocity is small compared with the longitudinal 2-velocity $\mathbf{U} = (U_1, U_2)$, $|U_3| \ll |\mathbf{U}|$, and it is neglected in the subsequent equations; (iii) the longitudinal velocity $\mathbf{U}$, the gravitational potential and the metric coefficients vary slightly along the liquid column.

By introducing the surface density $\sigma = \rho h$ and using \eqref{g11_g22}-\eqref{R_c}, we get shallow water equations on a stationary axisymmetric curved surface
\begin{equation}\label{EqShallowPhysCont}
\partial_t \sigma                                                       +
\frac{1}{r_0} \partial_l     \left( r_0\sigma U_{l}   \right) +
\frac{1}{r_0} \partial_\varphi \left( \sigma U_{\varphi}\right) =0.
\end{equation}
\begin{equation}\label{EqShallowPhysEuler_l}
(                         \partial_t           +
  U_l                    \partial_l            +
 \frac{U_{\varphi}}{r_0} \partial_\varphi)U_l  -
 \frac{r^{\prime}_0 }{r_0} U_{\varphi}^2       =
- \frac{1}{\sigma} \partial_l {\cal P}         - \phi_0^{\prime} ,
\end{equation}
\begin{equation}\label{EqShallowPhysEuler_phi}
 (\partial_t                                   +
  U_l \partial_l                               +
 \frac{U_{\varphi}}{r_0}\partial_\varphi)U_{\varphi}  +
  \frac{r^{\prime}_0}{r_0} U_l U_\varphi       =
- \frac{1}{\sigma r_0}\partial_\varphi {\cal P}.
\end{equation}
Integrated over the height of the liquid column pressure is defined as
\begin{equation}\label{cal_P}
  {\cal P} =  \frac{1}{2\rho} A(\mathbf{U}, l) \sigma^2,
\end{equation}
\begin{equation}\label{A}
  A(\mathbf{U}, l) =
   g_{\bot} +  U_{\varphi}^2 \sqrt{ 1-(r_0^{\prime})^2 }/{r_0}   - \theta^+_- {U_l^2}/{R_c} .
\end{equation}
The variable $g_{\bot}(l) = {\partial\phi}/{\partial n}|_{n=0}=-\mathbf{g}\cdot\mathbf{n} = g r_0^{\prime}$  is a component of the gravitational acceleration normal to the bottom. The shallow water is accelerated by the gravitational force with an {\it effective centrosymmetrical two-dimensional potential} $\phi_0(l) = gz_0(l) +{\rm const}$. In the limit of a funnel, deformable in plane ($r^{\prime}_0=1$) or a vertical tube ($r^{\prime}_0=0$), we have the standard equations of hydrodynamics on the plane and in the tube in cylindrical coordinates.

The features of the resulting equations are as follows:

(1) The pressure \eqref{cal_P} is proportional to the specific gravity of the liquid column \eqref{A}; if the liquid column in shallow water has no weight, there is no mechanism of propagation of long gravity waves.

(2) The shallow water on a curved bottom is not a barotropic medium. Its ``thermodynamics'' is determined not only by its internal state, but also by the state of its motion, and explicitly depends on the coordinates through the bottom's shape. In particular cases of horizontal or inclined straight bottom in the absence of the rotational speed the parameter A is constant and the fluid becomes barotropic.

(3) At the point at which the total centrifugal force becomes comparable to the gravity force, the pressure of the shallow water \eqref{cal_P},\eqref{A} becomes zero. At this point the flow separates from the bottom, and this point must be considered as a terminal to the shallow water flow. The flow separation can occur markedly above the lower edge of the funnel.

(4) Small perturbations with a wavelength $\lambda \gg h$ in shallow water propagate like an ordinary sound as dispersionless waves. However, the linear analysis of the equations \eqref{EqShallowPhysCont}-\eqref{A} shows that propagation of the gravity waves in shallow water with a curved bottom surface is anisotropic: wave velocities downstream and upstream relative to the moving fluid $c_{\pm} =  A_{,u}\sigma/(4\rho) \pm (A_{,u}^2\sigma^2/(16\rho^2) + A \sigma/\rho)^{1/2}$ do not coincide with each other in the absolute value. The difference in velocities is caused by the effect when downstream wave increases the magnitude of the centrifugal force repelling, reducing the weight of the liquid column, and the wave moving upstream, by contrast, increases the weight. The effect of anisotropy has the order of smallness $\sim h/R_c$ and in real situations($h/R_c \lesssim 1/10$) it will be barely discernable.

(5) Shock transitions in shallow water are possible as dissipationless hydraulic jumps. At the shock transition the density and the component of the velocity $U_{\bot}$ normal to jump undergo discontinuous change while the tangential component $U_{\parallel}$ remains the continuous one. On a curved bottom the value of $A$ may jump.

\section{Simulation of Newtonian and pseudo-Newtonian potentials. Gravitational length}

Let us choose such a funnel profile $(r_0(l), z_0(l))$ that the shallow water flow corresponds exactly to the accretion in the Newtonian, or more generally, in the pseudo-Newtonian potential of the type $\phi_{PN}=-GM/(l-l_g)$ \cite{Abra09}. Here $G$ is the gravity constant, $M$ and $l_g$ are the scale equivalents of the gravitating mass and the gravitational radius, respectively \footnote{Such a pseudo-potential correctly reproduces the effect of the loss of stability of circular orbits at $r < 3 r_g$.}). In the case of the Newtonian potential the parameter $l_g$ should be set equal to zero.

To drain the shallow water without waterfalls it is necessary that the dependence $r_0(z_0)$ be unambiguous and non-decreasing one: $dr_0/dz_0\ge 0$. Then the dependence $l(z_0)$ is also unambiguous and monotonous, and it can be defined by the formula
\begin{equation}\label{l(z)def}
l(z_0) = \int_{z_{out}}^{z_0} \sqrt{ 1+\left(d r_0/d z\right)^2 } dz + l_{out}.
\end{equation}
For the pseudo-Newtonian approximation of the potential dependence it is necessary to select $l(z_0)$ in such a way as to satisfy the condition
\begin{equation}\label{Eqphi}
\phi_0(l)\equiv  - GM/(l-l_g) = g(z_0 - z_{max}),
\end{equation}
where $z_{max}$ is the maximum height of the bottom level. The level $z_0=z_{max}$ is achieved asymptotically at $l\to\infty$. Actually one has to use $z_{in} \le z_{max}$ as an outer boundary in the setup (Fig.~\ref{fig:1}).

From \eqref{l(z)def}, \eqref{Eqphi} it follows that the spatial structure of the solution is determined by the scale
\begin{equation}\label{hg}
h_g = \left( GM/g \right)^{1/2},
\end{equation}
which we call {\it gravitational length}. The gravitational length for a point mass $M$ in the uniform gravitational field of a given intensity characterizes such a scale in which the depths of the potential gravitational well created by the mass $M$ and the external uniform field are equal. For comparison, the gravitational length $h_g=5$ cm from the surface of Earth corresponds to a body of mass $M\approx 367$ kilotons.

Considering together \eqref{l(z)def} and \eqref{Eqphi}, we find the solution in quadratures
\begin{equation}\label{l(z)Coul}
r_0(z_0) = \int_{z_{out}}^{z_0} \sqrt{ \left(h_g /(z_{max}-z)\right)^4 - 1 }\,\, dz + r_{out}.
\end{equation}
An explicit analytical expression for the integral written through the elliptic integral of the second kind is omitted because of its awkwardness. The funnel profile shown in Fig.~\ref{fig:1} corresponds exactly to \eqref{l(z)Coul}.

The solution \eqref{l(z)Coul} shows that the setup can not have a height $z_{max}-z_{out}$, greater than $h_g$. The tilting angles of the bottom at the boundary points are maximum permissible: when $z_0\to z_{max}$ the bottom is asymptotically horizontal one and at $z_0=z_{out}=z_{max}-h_g$ it has a vertical slope.

To recover the metric coefficients in equations \eqref{R_c}, \eqref{EqShallowPhysCont}, \eqref{EqShallowPhysEuler_phi}, \eqref{A} one should also have the dependence $r_0(l)$ which can be found from \eqref{l(z)def} and \eqref{l(z)Coul}. From there we also find $l_{out}=l_g+h_g$:
\begin{equation}\label{r(l)Coul}
r_0(l) = \int_{l_g+h_g}^l
\sqrt{ 1  - \left( h_g/(l-l_g)\right)^4 } dl + r_{out},
\end{equation}
and $z_0(l)$, which is obtained from \eqref{Eqphi}.

Substituting the potential from \eqref{Eqphi} in equation \eqref{EqShallowPhysEuler_l} for the stationary axisymmetric case in the absence of pressure and integrating over $l$, we find the energy integral for a particle radially falling into the funnel, which, as required, at $l_g=0$, coincides with equation \eqref{Int_energy_Sch} up to the change of notation:
\begin{equation}\label{Int_energy_SemKov}
  \frac{U_l^2}{2v_g^2} - \frac{h_g}{2l} = 0.
\end{equation}
Here the parabolic velocity $v_g=\sqrt{g h_g}$ fulfils the function of light velocity, the pseudo-radius $l$ fulfils the role of radius $R$, and the gravitational length $h_g$ serves as  the gravitational radius $r_g$.

The analogy with the spatial part of the Schwarzschild metric will become even more obvious if we rewrite the metric \eqref{g11_g22} of the space $V_2$ in the coordinates $(r_0, \varphi)$, taking into account \eqref{r(l)Coul}:
\begin{equation}\label{g_r0r0}
g_{\alpha\beta}  = {\rm diag}\left( (1-h_g^4/(l(r_0)-l_g)^4)^{-1}, r_0^2 \right).
\end{equation}
We see that $g_{\alpha\beta}$ has a removable singularity ($g_{r_0 r_0}\to\infty$) at $l(r_0)-l_g\to h_g$, and in this sense the gravitational length is a direct analogue of the gravitational radius in a laboratory model. However, the analogy is limited only by the spatial part of the metric. If we supplement the space $V_2$ by the time coordinate, then the time component $g_{tt}$ in the composite $(2+1)$-metric, unlike the Schwarzschild metric, will have no singularity at $r_0\to r_{out}$, since $g_{tt}\approx 1+2\phi_0/c^2$ (see \cite{LandLifsh}) and the gravitational potential on a laboratory scale varies insignificantly: $2\Delta\phi_0/c^2\sim 10^{-17}$ when $h_g = 5$ cm.

The effects which are the most interesting ones in the quasi-spherical accretion flow are observed either near the event horizon or in the vicinity of the accretion radius (the transition through the speed of sound, the possibility of formation of shock waves). Based on the linear dimensions of the setup which is characterized by parameters $z_{in}$, $h_{in}=h(z_{in})$, $h_g$, $r_{out}$ and $l_g$, we can estimate the equivalent parameters of accretion, the simulated mass of the black hole and the sound velocity in the accretion flow, the size of the simulated area in terms of gravitational accretion or radii. The latter one can be defined as
\begin{equation}\label{l_a}
l_a = GM/c^2_{s,in} = GM/g h_{in} = h_g^2/h_{in}.
\end{equation}
Taking for definiteness $h_{in} = 0.5$ cm and $h_g = 5$ cm, we obtain $l_a = 50$ cm, which allows us to estimate the setup diameter as $\sim 1.5 - 2$ m.

Among the important effects that we can simulate experimentally with shallow water is the asphericity of the accretion flow arising, for example, when the black hole moves relative to the medium (the so-called Bondi-Hoyle accretion) \cite{Beskin06}. Adding dipole component to the velocity field must make the flow unstable \cite{Ruffert95}. The laboratory modeling of such a flow is possible if the fluid is fed into the funnel nonradially, with a predetermined angle to a radius, depending on the azimuthal angle $\varphi$. Another important effect is the influence of the centrifugal barrier on the flow structure that can lead to the formation of double shock-wave transition in a thick accretion disk \cite{Chakra89}. This effect can be simulated by setting the fluid's rotation.

The generalities and distinctions in kind between the laboratory and theoretical quasi-classical models of quasi-spherical adiabatic accretion are summarized in Table~1.

\begin{landscape}
\hfill
 \begin{table*}%[H] add [H] placement to break table across pages
%\begin{center}
 \begin{tabular}{|c|c|c|}
\hline
%  % after \\: \hline or \cline{col1-col2} \cline{col3-col4} ...
                & \textrm{Theoretical model, $r\gg r_g$,}  &  Laboratory model\\
                & \textrm{quasi-classical description} & \\
\hline
  Metric & Euclidean   & non-Euclidean (pseudo-Schwarzschild) \\
  Geometry of flow on average
            & spherical or cylindrical & polar \\
  Gravitational potential & Newtonian (pseudo-Newtonian) & Newtonian (pseudo-Newtonian) \\
  Singularity of metric/potential & yes ($r_g\ne0$) &  yes ($h_g\ne0$) \\
  Relativistic kinematics & insignificant &  no \\
  Compressibility & yes & yes \\
  Medium & barotropic & non-barotropic \\
  Adiabatic index & $\gamma\le 5/3$ & $\gamma=2$ \\
  Shock jumps & adiabatic, dissipative & adiabatic, dissipationless \\
  Acoustic horizon & yes & yes \\
   Degree of cumulation & $\sim 10\div 10^{11}$  $\{\sim r_a/r_g\}$ & $\sim 10\div 100$  $\{\sim r_{in}/r_{out}\}$ \\
  \hline
 \end{tabular}
%\end{center}
 \caption{\label{c1}
Comparison of flow peculiarities for the theoretical and laboratory models of quasi-spherical adiabatic accretion.}
 \end{table*}
\hfill
\end{landscape}

\bigskip\medskip\noindent {\bf Acknowledgements}
\medskip
A.V.S. thanks Russian Foundation for Basic Research for the financial support of the project 16-31-00466-mol\_a.

\end{document}